\documentclass[proceedings]{JHEP}
\usepackage{cite,epsfig}

\newbox\mybox
         
\newcommand\fverb{\setbox\mybox=\hbox\bgroup\verb}
\newcommand\fverbdo{\egroup\medskip\noindent\fbox{\unhbox\mybox}\ }
\newcommand\fverbit{\egroup\item[\fbox{\unhbox\mybox}]}

\font\beeg=cmr17 scaled 1600		
\newcommand\init[1]{\setbox\mybox=\hbox{{\beeg #1}~}%
		   \noindent\global\hangindent=\wd\mybox\global\hangafter-2%
		   \sc\smash{\llap {\lower 13.2pt \box\mybox}}}

\newcommand{\nn}{\nonumber}
\newcommand{\ve}{\varepsilon}
\newcommand{\IM}{\mbox{\rm Im}}
\newcommand{\RE}{\mbox{\rm Re}}
\newcommand{\eqn}[1]{(\ref{#1})}
\newcommand{\mev}{\mbox{\rm MeV}}
\newcommand{\gev}{\mbox{\rm GeV}}
\newcommand{\epe}{\varepsilon'/\varepsilon}

\title{\boldmath Theoretical status of $\epe$ 
\hfill {\small HD-THEP-99-51}
\unboldmath}

\author{
Matthias Jamin\thanks{Heisenberg fellow.}\\
Institut f\"ur Theoretische Physik\\
Philosophenweg 16, 69120 Heidelberg, Germany\\
E-mail: \email{M.Jamin@ThPhys.Uni-Heidelberg.DE}}

\conference{Heavy Flavours 8, Southampton, UK, 1999}

\abstract{
The present theoretical status of the parameter for {\em direct} CP violation
$\epe$ in the Standard Model is reviewed and compared with most recent
experimental measurements of the same quantity. After collecting the basic
expressions for $\epe$, the situation of hadronic matrix element calculations
is summarised. The matrix elements constitute the dominant source of
uncertainty for theoretical determinations of $\epe$. For central values of
the input parameters, the numerical analysis then yields results which are
generally below the experimental data. Possible reasons for these findings
are discussed.}

\begin{document} 

{\init This year} has experienced a revived strong interest in the parameter
$\epe$ which quantifies {\em direct} CP violation in the neutral kaon
system, due to the recent experimental measurements by the KTeV and NA48
collaborations:
\begin{eqnarray}
\RE\,(\epe) & = & (28.0\pm 4.1) \cdot 10^{-4} \;\;
\hbox{KTeV \cite{KTeV}} \,, \nn \\
\RE\,(\epe) & = & (18.5\pm 7.3) \cdot 10^{-4} \;\;
\hbox{NA48 \cite{NA48}} \,. \nn
\end{eqnarray}
Both measurements confirm the large value found previously by the NA31
collaboration \cite{NA31} and give rise to the new world average \cite{ago:99}
\begin{equation}
\RE\,(\epe) = (21.4\pm 4.0) \times 10^{-4} \,.
\end{equation}
Experimentally, the signature for direct CP violation is a deviation of
the ratio $|\eta_{+-}/\eta_{00}|^{\,2}$ from unity, where
\begin{eqnarray}
\eta_{+-} & \equiv & \frac{A(K_L\rightarrow\pi^+\pi^-)}
                          {A(K_S\rightarrow\pi^+\pi^-)} \,, \\
\eta_{00} & \equiv & \frac{A(K_L\rightarrow\pi^0\pi^0)}
                          {A(K_S\rightarrow\pi^0\pi^0)} \,.
\end{eqnarray}

Theoretically, it is more convenient to consider quantities were the
final state pions are in a definite isospin state:
\begin{eqnarray}
\ve    & \equiv & \frac{A(K_L\rightarrow(\pi\pi)_{I=0})}
                       {A(K_S\rightarrow(\pi\pi)_{I=0})} \,, \\
\omega & \equiv & \frac{A(K_S\rightarrow(\pi\pi)_{I=2})}
                       {A(K_S\rightarrow(\pi\pi)_{I=0})} \,.
\end{eqnarray}
The parameter $\ve$ measures {\em indirect} CP violation in the neutral
kaon system which displays the fact that the physical mass eigenstates
$K_L$ and $K_S$ are not eigenstates of CP, but have small admixtures of
the order of $\ve$ from the opposite CP parity. The second parameter $\omega$
has nothing to do with CP violation, but is introduced for convenience.
It parametrises the so called $\Delta I=1/2$ rule which states that the
isospin zero final state is much enhanced over the isospin two final state.
In other words, the $\Delta I=1/2$ transition is strongly enhanced over
the $\Delta I=3/2$ transition.

With the help of these quantities, the parameter $\ve'$ can be defined as
\begin{equation}
\ve' \equiv \frac{1}{\sqrt{2}}\biggl[\,
\frac{A(K_L\rightarrow(\pi\pi)_{I=2})}{A(K_S\rightarrow(\pi\pi)_{I=0})}
-\ve\cdot\omega\,\biggr] \,.
\end{equation}
Using the isospin decomposition of the final two-pion state it is a
simple exercise to find the relation between the experimental parameters
$\eta_{+-}$, $\eta_{00}$ and the theoretical parameters $\ve$, $\ve'$ and
$\omega$:
\begin{eqnarray}
\label{etapm}
\eta_{+-} & = & \ve + \frac{\ve'}{1+\omega/\sqrt{2}} \,, \\
\label{eta00}
\eta_{00} & = & \ve - \frac{2\,\ve'}{1-\sqrt{2}\,\omega} \,.
\end{eqnarray}
From these relations and the measured results for $\eta_{+-}$ and
$\eta_{00}$ \cite{pdg} one can deduce an experimental value for $\ve$:
\begin{eqnarray}
\ve =  (2.280 \pm 0.013) \cdot 10^{-3} \, e^{i(43.5\pm0.1)^o} \,.
\end{eqnarray}
Employing CPT invariance in addition, the phase has been extracted from
the known $K_L$-$K_S$ mass difference.

To proceed further, one expresses the parameters $\ve$, $\ve'$ and
$\omega$ in terms of isospin amplitudes $A_I$ and final state interaction
phase shifts $\delta_I$ which are defined by
\begin{eqnarray}
\label{AI}
A(K^0\rightarrow(\pi\pi)_I) & = & i\,A_I\,e^{i\delta_I} \,, \\
A(\bar K^0\rightarrow(\pi\pi)_I) & = & -\,i\,A_I^*\,e^{i\delta_I} \,.
\end{eqnarray}
If CP were conserved, the amplitudes $A_I$ would be real. Making use of the
relation between the strangeness eigenstates $K^0$, $\bar K^0$ and
the mass eigenstates $K_L$, $K_S$ (further details can for example be
found in ref. \cite{deRaf}) the parameter $\ve'$ can be written as
\begin{equation}
\ve' = \frac{1}{\sqrt{2}}\,(1-\ve^2)\,\IM\biggl(\frac{A_2}{A_0}\biggr)
e^{i(\pi/2+\delta_2-\delta_0)} \,,
\end{equation}
or expressed in a different, equivalent form:
\begin{equation}
\label{epe1}
\ve' = \frac{(1-\ve^2)e^{i(\pi/2+\delta_2-\delta_0)}}{\sqrt{2}\,
(1+\xi_0^2)\,\RE A_0}\biggl[\,\IM A_2 - \xi_0\,\RE A_2\,\biggr] \,,
\end{equation}
where $\xi_0\equiv\IM A_0/\RE A_0$. So far no approximations have been
performed and all equations are exact. In particular, equation~\eqn{epe1}
will be the starting point for our theoretical analysis of the ratio $\epe$.

Analogously, to a very good approximation the parameter $\omega$ can
be expressed as
\begin{equation}
\label{omega}
\omega \approx \frac{\RE A_2}{\RE A_0} \, e^{i(\delta_2-\delta_0)} =
\frac{1}{22.2}\,e^{-i(45\pm 6)^o} \,,
\end{equation}
again reflecting the $\Delta I=1/2$ rule. The difference of the strong
interaction phase shifts has been obtained in ref. \cite{gm:91}. A
consequence of the $\Delta I=1/2$ rule is that the isospin zero contribution
to $\ve'$ is suppressed by the small quantity $\omega$ compared to the
isospin two component. Phenomenological implications of this observation
will be further discussed below.

Using the relations \eqn{etapm} and \eqn{eta00} together with the
experimental result \eqn{omega} for $\omega$, up to corrections of
the order of 2\% one finds
\begin{equation}
\RE\biggl(\frac{\ve'}{\ve}\biggr) \approx \frac{1}{6}\,\biggl[\,
\left|\frac{\eta_{+-}}{\eta_{00}}\right|^2 - 1\,\biggr] \,.
\end{equation}
At present, these corrections are still much below the experimental
uncertainties and can be safely neglected.

\section{\boldmath Basic formulae for $\epe$\unboldmath}

Neglecting the tiny corrections of order $\ve^2$ and $\xi_0^2$ in
equation~\eqn{epe1}, the central expression for $\epe$ takes the form
\begin{equation}
\label{epe2}
\frac{\ve'}{\ve} \approx \frac{e^{i(\pi/4+\delta_2-\delta_0)}}{\sqrt{2}\,
|\ve|\,\RE A_0}\biggl[\,\IM A_2 - |\omega|\,\IM A_0\,\biggr] \,.
\end{equation}
Since it was found that the phase shift difference $\delta_0-\delta_2
\approx\pi/4$, within the uncertainties $\epe$ turns out to be real.
Calculating the amplitudes $A_I$ in the framework of the operator
product expansion and applying the renormalisation group equation,
the basic formulae for $\epe$ is found to be
\begin{equation}
\label{epe3}
\frac{\ve'}{\ve} = \IM\lambda_t \Big[\,P^{(1/2)}-P^{(3/2)}\,\Big] \,.
\end{equation}
Here $\lambda_t\equiv V_{td}V_{ts}^*$ with $V_{ij}$ being the elements of the
quark mixing or Cabibbo-Kobayashi-Maskawa (CKM) matrix. To an excellent
approximation one has
\begin{equation}
\IM\lambda_t \approx |V_{ub}||V_{cb}| \sin\delta
\end{equation}
with $\delta$ being the CP-violating phase in the standard parametrisation
of the CKM matrix \cite{pdg}.

Further, the $P^{(\Delta I)}$ are given by
\begin{eqnarray}
P^{(1/2)} & = & r \sum\limits_i y_i(\mu) \langle(\pi\pi)_0|Q_i(\mu)|K\rangle
(1-\Omega_{IB}) \,, \nn \\
P^{(3/2)} & = & \frac{r}{|\omega|} \sum\limits_i y_i(\mu) \langle(\pi\pi)_2|
Q_i(\mu)|K\rangle  \,,
\end{eqnarray}
with
\begin{equation}
\label{rval}
r = \frac{G_F|\omega|}{2|\ve|\RE A_0} = 346\,\gev^{-3} \,.
\end{equation}
The $y_i(\mu)$ are Wilson coefficient functions corresponding to the
operators $O_i$ and $\mu$ denotes the renormalisation scale which for our
analysis will be of order $1\,\gev$. In addition to the renormalisation
scale dependence, both, the Wilson coefficients $y_i$, and the hadronic matrix
elements of operators $O_i$, depend on the renormalisation scheme. Of course,
up to the calculated order, for the physical quantities $P^{(\Delta I)}$ these
dependencies should cancel. Present-day values for the Wilson coefficients
$y_i(m_c)$ for two commonly used schemes can be found in ref.~\cite{epe99}.
Finally, $\Omega_{IB}$ is an isospin breaking correction which arises because
$m_u\neq m_d$. In the numerical analysis we use $\Omega_{IB}=0.25\pm 0.08$
\cite{omIB1,omIB2} but we shall further comment on isospin breaking effects
below. 

Omitting negligible contributions from dimension five magnetic-dipole
operators, the leading contribution to $\epe$ results from dimension six
four-quark operators. These can be classified into $Q_{1,2}$ (current-current),
$Q_{3-6}$ (QCD penguin) and $Q_{7-10}$ (electroweak penguin) according
to the type of Feynman diagrams from which they arise. Explicit
expressions for all operators can for example be found in ref. \cite{bjl:93b}
and the review article \cite{bbl:96}. Here we only give the two dominant
QCD and electroweak penguin operators:
\begin{eqnarray}
Q_6 & = & (\bar s_\alpha  d_\beta)_{V-A}\!\!\sum\limits_{q=u,d,s}
          (\bar q_\beta \,q_\alpha)_{V+A} \,, \\
Q_8 & = & \frac{3}{2} \,(\bar s_\alpha  d_\beta)_{V-A}\!\!\sum\limits_{q=u,d,s}
          \!e_q (\bar q_\beta  \,q_\alpha)_{V+A} \,,
\end{eqnarray}
where $\alpha$, $\beta$ are colour indices and $e_q$ denotes the electric
quark charges. Although the contribution of the electroweak penguin $Q_8$
is suppressed by a factor $\alpha_{em}$ it is enhanced by $1/|\omega|$
and thus, as we shall see further below, has some impact on the value
of $\epe$.

\section{\boldmath History of $\epe$ calculations\unboldmath}

To be able to appreciate the achievements of the $\epe$ calculations in
the last almost 25 years since the first estimate of $\epe$ \cite{egn:76},
let me briefly review their history.

The first estimate of $\epe$ \cite{egn:76} assumed $m_t\ll M_W$, only included
QCD penguins which were introduced first in ref.~\cite{vzs:75} and
omitted renormalisation group effects. Nevertheless, just by chance the
resulting value $1/450$ is surprisingly close to the current world
average. Renormalisation group effects in the leading logarithmic
approximation have first been taken into account in \cite{gw:79}. For
$m_t \ll M_W$ only QCD penguins play a substantial role. First extensive
phenomenological analyses in this approximation can be found in \cite{bss:84}.

Over the eighties these calculations were refined through the inclusion of
QED penguin effects for $m_t \ll M_W$ \cite{omIB1,bw:84}, the inclusion of
isospin breaking in the quark masses \cite{omIB1,omIB2} and through improved
estimates of hadronic matrix elements in the framework of the $1/N_c$
approach \cite{bbg:87}. This era of $\epe$ culminated in the analyses in
\cite{fr:89,bbh:90}, where QCD penguins, electroweak penguins and the
relevant box diagrams were included for arbitrary top quark masses. The
strong cancellation between QCD penguins and electroweak penguins for
$m_t > 150~\gev$ found in these papers was later confirmed by other authors
\cite{pw:91}.

During the nineties considerable progress has been achieved by calculating
complete next-to-leading order (NLO) corrections to the Wilson coefficients
$y_i(\mu)$ \cite{bjl:93b,bjlw:92,bjl:93a,rome:93}. Down to a scale of order
$1\,\gev$ the corrections turned out to be modest which allows good control
over the short distance part in the operator product expansion. Together
with the NLO corrections to $\ve$ and $B^0$--$\bar B^0$ mixing
\cite{bjw:90,hn:94,ukjs:98}, this allowed for an improved NLO analysis of
$\epe$ including constraints from the observed indirect CP violation ($\ve$)
and $B_{d,s}^0$--$\bar B_{d,s}^0$ mixings ($\Delta M_{d,s}$). Progress in the
determination of the $V_{ub}$ and $V_{cb}$ elements of the CKM matrix and
in particular the determination of the top quark mass $m_t$ had of course
also an important impact on $\epe$.

Nevertheless, it is fair to say that calculations of the long distance
part, the hadronic matrix elements, have not yet reached a level which would
match the NLO calculations of the Wilson coefficients. Long distance
physics inevitably involves confinement effects and thus non-perturba-tive
methods are required for the calculation of the matrix elements. In principal
lattice QCD calculations should be able to give matrix elements with the
correct scale and scheme dependencies to match the coefficient functions,
but the most important matrix element of $Q_6$ has so far not been obtained
reliably.

Other methods which are based
on effective theories like chiral perturbation theory (ChPT) or the
$1/N_c$ expansion suffer from problems because a sound matching to the
Wilson coefficients with the correct scale and especially scheme
dependencies is not obvious. Finally, QCD sum rules could be useful, but
except for the $K^0$--$\bar K^0$ mixing parameter $B_K$, the calculation
of hadronic matrix elements in this approach has not been developed far
enough to be competitive to the other methods. In the next section we
shall thus summarise the status of the hadronic matrix elements relevant
for $\epe$ concentrating on the dominant contributions $\langle Q_6\rangle_0$
and $\langle Q_8\rangle_2$.

\section{Hadronic matrix elements}

For a discussion of the hadronic matrix elements it is convenient to
introduce the so called $B$-para-meters which quantify the deviation of
the full matrix elements to the vacuum-saturation or factorisation
approximation were the four-quark operator is factorised in the product
of two currents by inserting a vacuum intermediate state. This
approximation usually serves as a first estimate of hadronic matrix elements.
Thus we define
\begin{eqnarray}
\langle Q_6\rangle_0 & \equiv & B_6^{(1/2)} \langle Q_6\rangle_0^{(vac)} \,, \\
\langle Q_8\rangle_2 & \equiv & B_8^{(3/2)} \langle Q_8\rangle_2^{(vac)} \,,
\end{eqnarray}
and the factorisation approximation corresponds to $B_6^{(1/2)}=B_8^{(3/2)}=1$.
These values even hold in the large-$N_c$ limit \cite{bbg:87} which means
that in this limit factorisation is exact.

Because it turns out that the factorised matrix elements
$\langle Q_6\rangle_0^{(vac)}$ and $\langle Q_8\rangle_2^{(vac)}$ are
proportional to $1/m_s^2$, the value of the strange quark mass enters the
analysis of $\epe$. This is not necessary as for example on the lattice
the matrix elements $\langle Q_6\rangle_0$ and $\langle Q_8\rangle_2$
are calculated directly and no direct dependence on the strange mass arises.
However, although the matrix elements depend on the renormalisation
scale, a careful analysis \cite{bjl:93b} showed that this dependence is
almost completely covered by the scale dependence of the strange mass
and $B_6^{(1/2)}$, $B_8^{(3/2)}$ for energies of interest are practically
scale independent. In fact, this statement is exact in the large-$N_c$ limit
since in this limit the anomalous dimensions of $Q_6$ and $Q_8$ are minus
twice the mass anomalous dimension \cite{bjl:93b}. For this reason we stick
to the discussion of the $B$-parameters which make a comparison of different
methods which work at different scales easier.

The status of strange quark mass determinations has been recently
summarised in refs. \cite{epe99,bur:99,mar:99}. For further references the
reader is referred to these works. Most precise values of the strange quark
mass come from lattice QCD and QCD sum rule calculations. As a present
average, we quote
\begin{equation}
\label{ms2}
m_s(2\,\gev) = (110\pm 20)\,\mev \,.
\end{equation}
Unquenched lattice calculations yield somewhat smaller values but at present
the information is not precise enough to be conclusive. In the approach to
the analysis of $\epe$ of ref.~\cite{bjl:93b} it is convenient to calculate
the matrix elements at the scale $m_c$ because at that scale many of the
remaining hadronic matrix elements can be determined from CP-conserving
$K\rightarrow\pi\pi$ decays. Thus we also present the value of the strange
mass at that scale:
\begin{equation}
\label{msmc}
m_s(m_c) = (130\pm 25)\,\mev \,,
\end{equation}
where $m_c=1.3\,\gev$ has been used.

\TABULAR[ht]{|c|c|c|}
{\hline
Method & $B_6^{(1/2)}$ & $B_8^{(3/2)}$ \\
\hline
Lattice \cite{Blat} & -- & 0.69 -- 1.06 \\
Large-$N_c$ \cite{BNc} & 0.72 -- 1.10 & 0.42 -- 0.64 \\
ChQM \cite{BChQM} & 1.07 -- 1.58 & 0.75 -- 0.79 \\
\hline}
{\label{B6B8} Results for $B_6^{(1/2)}$ and $B_8^{(3/2)}$ obtained in
different approaches.}

Values for $B_6^{(1/2)}$ and $B_8^{(3/2)}$ obtained in various approaches
are collected in table~\ref{B6B8}. The lattice results have been calculated
at $\mu=2\,\gev$. Concerning the lattice results for $B_6^{(1/2)}$, old
calculations gave values around one with errors of the order of 30\%.
However, a recent work \cite{pk:98} shows that NLO QCD corrections in the
relation between lattice and continuum operators are so huge that at present
there is no solid prediction for $B_6^{(1/2)}$ on the lattice. The situation
of $B_8^{(3/2)}$ is better. Here most approaches find a suppression of
$B_8^{(3/2)}$ below unity by roughly 20\%. Further discussion of the lattice
approach and additional references can be found in \cite{cfglm:99}.

The average value of $B_6^{(1/2)}$ in the large-$N_c$ approach including
full order $p^2$ and $p^0/N_c$ contributions as given in table~\ref{B6B8}
is close to $1$ whereas the suppression of $B_8^{(3/2)}$ compared to the
large-$N_c$ limit is stronger than on the lattice. The uncertainty comes
from a variation of the cut-off scale $\Lambda$ in the effective theory.
On the other hand it has been found \cite{hkps:99} that a higher order
term ${\cal O}(p^2/N_c)$ enhances $B_6^{(1/2)}$ to 1.6. This result is
clearly interesting. Yet, in view of the fact that other $p^2/N_c$ terms
as well as $p^4$ and $p^0/N_c^2$ terms have not been calculated, it is
premature to take this enhancement seriously.

Finally, the chiral quark model (ChQM) gives values for $B_6^{(1/2)}$ as
high as $1.33\pm 0.25$. On the other hand, $B_8^{(3/2)}$ in this approach is
well compatible with the lattice and large-$N_c$ calculations. Guided by the
results presented above and biased to some extent by the results from the
lattice and large-$N_c$ approach to hadronic matrix elements, the status
of determinations of $B_6^{(1/2)}$ and $B_8^{(3/2)}$ is summarised as:
\begin{equation}
B_6^{(1/2)} = 1.0 \pm 0.3 \,, \quad
B_8^{(3/2)} = 0.8 \pm 0.2 \,. \quad
\end{equation}
In addition, in our numerical analysis we shall always keep
$B_6^{(1/2)}\geq B_8^{(3/2)}$ as it is found with all non-perturbative
methods.

\section{\boldmath Numerical analysis of $\epe$\unboldmath}

Being one of the authors of a recent analysis of $\epe$ \cite{epe99} by the
so-called {\it Munich group}, I take the liberty to mainly concentrate on
this work but I shall comment on other recent analyses below.

\FIGURE{
\epsfig{file=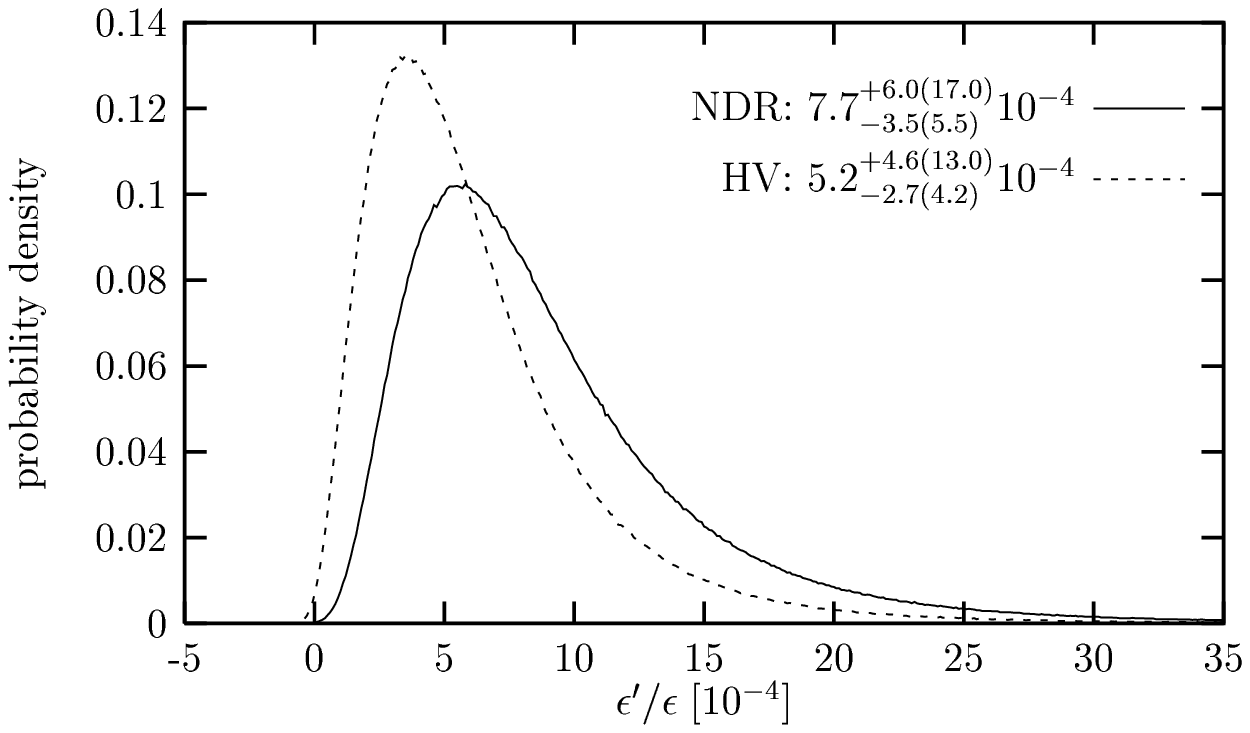,width=15cm}
\caption{Probability distribution for $\epe$ in the NDR and HV schemes.}
\label{fig:epe}}

Before we can proceed with the calculation of $\epe$, we still have to
extract the value of $\sin\delta$, or equivalently, $\IM\lambda_t$ directly.
This can be obtained from a standard analysis of the unitarity triangle
which uses data for $|V_{cb}|$, $|V_{ub}|$, $\ve$, $\Delta M_d$ and
$\Delta M_s$, where the last two measure the size of
$B_{d,s}^0$--$\bar B_{d,s}^0$ mixing. This type of analysis is rather
well known and the reader is referred to the literature for details
\cite{cfglm:99,bjl:96,ut:99}. As our result for $\IM\lambda_t$, we quote
\cite{epe99}
\begin{equation}
\label{Imlat}
\IM\lambda_t = (1.33 \pm 0.14)\cdot 10^{-4} \,.
\end{equation}

At this point it is instructive to present a formula which is not to be
used for any serious analysis, nevertheless in a crude approximation
displays the dependence of $\epe$ on the most important parameters
collected in table~\ref{input}:
\begin{eqnarray}
\label{crude}
\frac{\ve'}{\ve} & \approx & 13\,\IM\lambda_t \biggl[\frac{130\,\mev}{m_s(m_c)}
\biggr]^2 \biggl[\, B_6^{(1/2)}(1-\Omega_{IB}) \nn \\
& & \hspace{-8mm} -\,0.4 B_8^{(3/2)}\Big(\frac{m_t(m_t)}{165\,\gev}\Big)^{2.5}
\biggr]\biggl(\frac{\Lambda_{\overline{{\rm MS}}}^{(4)}}{340\,\mev}\biggr) \,.
\end{eqnarray}
This formula exhibits clearly the dominant uncertainties which reside in
the values of $B_6^{(1/2)}$, $B_8^{(3/2)}$, $m_s$, $\Lambda_{\overline{{\rm MS}}}^{(4)}$ and $\Omega_{IB}$. Because of the rather accurate value of the top
quark mass, the resulting uncertainty in $\epe$ amounts only to a few percent.

Let us now continue with the full analysis.
Using equations~\eqn{epe3}--\eqn{rval} and \eqn{Imlat}, the values for the
Wilson coefficient functions $y_i$ \cite{epe99,bjl:93b}, the values of the
$B$-parameters and strange quark mass as discussed in the previous section,
expressions for the matrix elements in the vacuum insertion approximation
\cite{epe99,bjl:93b}, as well as the value of $\Omega_{IB}$ given in section~2,
we are in a position to calculate $\epe$.

\TABULAR[ht]{|c|c|c|}
{\hline
Quantity & Value & Reference \\
\hline
$\Lambda_{\overline{{\rm MS}}}^{(4)}$ & $(340 \pm 50)\,\mev$ & \cite{pdg} \\
$m_s(m_c)$ & $(130 \pm 25)\,\mev$ & See text \\
$m_t(m_t)$ & $(165 \pm 5)\,\gev$ & \cite{pdg} \\
$B_6^{(1/2)}$ & $1.0 \pm 0.3$ & See text \\
$B_8^{(3/2)}$ & $0.8 \pm 0.2$ & See text \\
\hline}
{\label{input} Collection of main input parameters.}

For an estimation of the uncertainties in the determination of $\epe$
we follow two different stra-tegies:

\begin{itemize}
\item Method 1: All experimental and theoretical input parameters are
scanned independently within their ranges to produce the minimal and
maximal value for $\epe$.

\item Method 2: A Monte Carlo analysis is performed were all experimental
input parameters are simulated with Gaussian errors and all theoretical
input parameters with flat errors. The result for $\epe$ is then extracted
from a statistical analysis of the resulting probability distribution.
\end{itemize}

In the so-called NDR scheme, our result for the scanning method is:
\begin{equation}
1.05\cdot 10^{-4} \leq \epe \leq 28.8 \cdot 10^{-4} \,.
\end{equation}
The values found in the HV scheme, a second scheme considered by us, are
generally 20-30\% lower. This reflects the fact that at present the scheme
dependence of the matrix elements is not fully under control and the
difference in the results is due to a residual scheme dependence.
For the statistical analysis in the NDR scheme, we obtain
\begin{equation}
\epe = (7.7^{+6.0}_{-3.5})\cdot 10^{-4} \,,
\end{equation}
were similar comments apply for the result in the HV scheme. We have quoted
the median and 68\% confidence intervals because the resulting distribution
is rather asymmetric. A plot of the corresponding probability distribution
is shown in figure~\ref{fig:epe}.

\TABLE{
\begin{tabular}{|c|c|c|c|c|c|}
\hline
 $B_6^{(1/2)}$ & $B_8^{(3/2)}$ & $m_s(m_c)[\mev]$ &
  Central & $\Lambda_{\overline{\rm MS}}^{(4)}=390\mev $ & Maximal \\
\hline
      &     & $105$ &  20.2 & 23.3 & 28.8\\
 $1.3$&$0.6$& $130$ &  12.8 & 14.8 & 18.3\\
      &     & $155$ &   8.5 &  9.9 & 12.3 \\
 \hline
      &     & $105$ &  18.1 & 20.8 & 26.0\\
 $1.3$&$0.8$ & $130$ & 11.3 & 13.1 & 16.4\\
      &     & $155$ &   7.5 &  8.7 & 10.9\\
 \hline
      &     & $105$ &   15.9 & 18.3 & 23.2\\
 $1.3$&$1.0$ & $130$ &  9.9 &  11.5 & 14.5\\
      &     & $155$ &   6.5  &  7.6 &  9.6\\
 \hline\hline
      &     & $105$ &   13.7 & 15.8 & 19.7\\
 $1.0$&$0.6$ & $130$ &  8.4 &  9.8& 12.2 \\
      &     & $155$ &   5.4 &  6.4 & 7.9 \\
 \hline
      &     & $105$ &   11.5 & 13.3 & 16.9\\
$1.0$&$0.8$ & $130$  &  7.0 &   8.1 & 10.4\\
     &     & $155$ &   4.4 &    5.2 &  6.6\\
 \hline
     &     & $105$ &   9.4 &   10.9 & 14.1 \\
$1.0$&$1.0$ & $130$  &  5.5 &   6.5 &  8.5 \\
     &     & $155$ &   3.3 &    4.0 &  5.2\\
\hline
\end{tabular}
\caption{Values of $\epe$ in units of $10^{-4}$ for specific values
of $B_6^{(1/2)}$, $B_8^{(3/2)}$, $m_s(m_c)$ and other parameters as
explained in the text.}
\label{epetab}}

In table~\ref{epetab} we show the values of $\epe$ in units of $10^{-4}$ for
specific values of $B_6^{(1/2)}$, $B_8^{(3/2)}$ and $m_s(m_c)$ as calculated
in the NDR scheme. The corresponding values in the HV scheme are lower as
discussed above. The fourth column shows the results for central values of
all remaining parameters. The comparison of the fourth and fifth column
demonstrates how $\epe$ is increased when $\Lambda_{\overline{\rm MS}}^{(4)}$
is raised from $340\,\mev$ to $390\,\mev$. As stated in equation~\eqn{crude},
$\epe$ is roughly proportional to $\Lambda_{\overline{\rm MS}}^{(4)}$.
Finally, in the last column maximal values of $\epe$ are given. To this end
we have scanned all parameters relevant for the analysis of $\IM\lambda_t$
within one standard deviation and have chosen
$\Lambda_{\overline{\rm MS}}^{(4)}=390\,\mev$. Comparison of the last two
columns demonstrates the impact of the increase of $\IM\lambda_t$ from its
central to its maximal value and the variation of $m_t$.

We observe that the most probable values for $\epe$ are in the ball park of
$10^{-3}$. On the other hand table~\ref{epetab} shows that for particular
choices of input parameters, values for $\epe$ as high as $(2-3)\cdot 10^{-3}$
cannot be excluded. The largest uncertainties reside in $m_s$, $B_6^{(1/2)}$
and $B_8^{(3/2)}$. $\epe$ increases by roughly a factor of $2.3$ when
$m_s(m_c)$ is changed from $155\,\mev$ to $105\,\mev$. The increase of
$B_6^{(1/2)}$ from $1.0$ to $1.3$ increases $\epe$ by approximately 60\%
whereas corresponding changes due to $B_8^{(3/2)}$ are around 40\%. The
combined uncertainty due to $\IM\lambda_t$ and $m_t$ are roughly 25\% and
the uncertainty coming from $\Lambda_{\overline{\rm MS}}^{(4)}$ amounts
to 15\%.

\FIGURE{
\epsfig{file=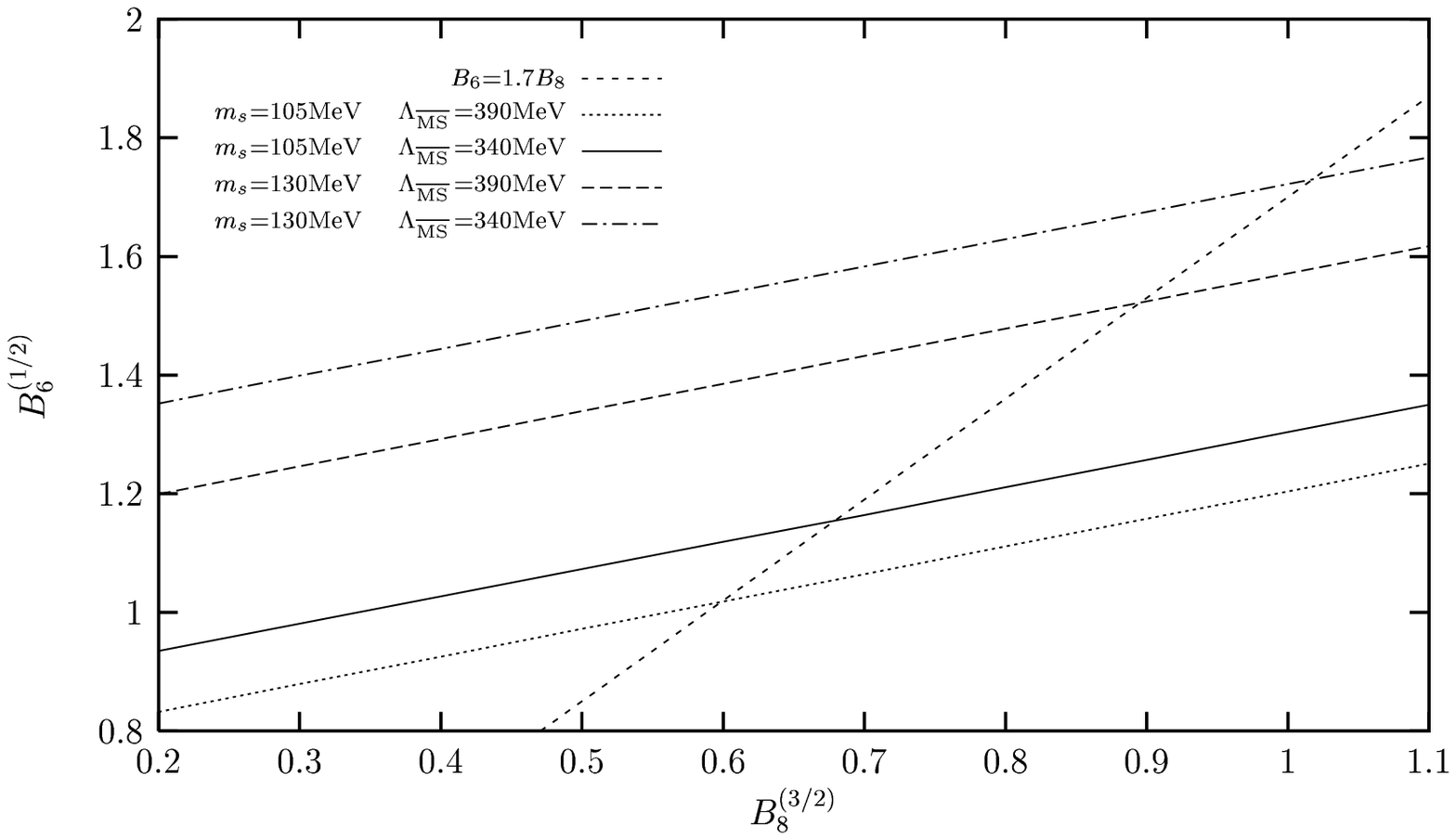,width=13cm}
\caption{Minimal value for $B_6^{(1/2)}$ consistent with
 $\epe\geq 2\cdot 10^{-3}$ as a function of $B_8^{(3/2)}$.}
\label{fig:b6b8}}

In figure~\ref{fig:b6b8}, we show the minimal value of $B_6^{(1/2)}$ for two
choices of $m_s(m_c)$ and $\Lambda_{\overline{\rm MS}}^{(4)}$ as a function
of $B_8^{(3/2)}$ for which the theoretical value of $\epe$ is higher than
$2\cdot 10^{-3}$. To obtain this plot we have varied all other parameters
within their uncertainties. We show also a line which corresponds to the
relation
\begin{equation}
B_6^{(1/2)} = 1.7 \cdot B_8^{(3/2)} \,.
\end{equation}
This relation holds in the large-$N_c$ approach to the hadronic matrix
elements independent of the cut-off scale $\Lambda$. One observes that as
long as $B_8^{(3/2)}\geq 0.6$, the parameter $B_6^{(1/2)}$ is required to
be larger than unity.

Let me now come to a comparison with other recent analyses of $\epe$.
All groups use the Wilson coefficient functions calculated in refs.
\cite{bjl:93b,bjlw:92,bjl:93a,rome:93}. Therefore, the differences in
$\epe$ result dominantly from different values for the hadronic matrix
elements and to some extent different input parameters needed for the
determination of $\IM\lambda_t$.

A very recent analysis by the Rome group \cite{cfglm:99}, besides
$B_6^{(1/2)}$ using matrix elements from the lattice in the HV scheme,
finds $\epe=(4-7)\cdot 10^{-4}$, completely compatible with the results
presented above in the statistical analysis. Also their scanning results
are similar except that the Rome group does not use the constraint
$B_6^{(1/2)}\geq B_8^{(3/2)}$ and allows for a larger error in $B_6^{(1/2)}$
which results in values as low as $\epe=-\,1\cdot 10^{-3}$.

Matrix elements in the large-$N_c$ approach were used in the analysis of
$\epe$ by the Dortmund group \cite{hkps:99}. With the exception of the
large correction of ${\cal O}(p^2/N_c)$ which was found to enhance
$B_6^{(1/2)}$ up to $1.6$, the matrix elements in the large-$N_c$ approach
are in agreement to the values used in our analysis. Thus, of course, also
the resulting values for $\epe$ agree and the larger value for
$B_6^{(1/2)}$ would bring $\epe$ much closer to the experimental average.

Finally, the Trieste group \cite{BChQM} generally finds higher values of
$\epe$, with the central value around $17\cdot 10^{-4}$ and consequently
consistent with the experimental findings. The main reason is a higher
value of $B_6^{(1/2)}$ as obtained from the chiral quark model. In principal
one could compare the results in the large-$N_c$ and ChQM approaches, but,
whereas the former was regularised with a cut-off, in the latter calculation
dimensional regularisation was used and a direct comparison is not possible.

\section{Discussion}

As the numerical analysis above shows, for present values of the theoretical
input parameters, estimates of $\epe$ in the Standard Model are typically
below the experimental data. However, as the scanning analysis demonstrates,
for suitably chosen parameters $\epe$ in the Standard Model can be made
consistent with the data. Yet, this only happens if several of the relevant
parameters are simultaneously close to extreme values of the ranges given
in table~\ref{input}. On the other hand also $B_6^{(1/2)}\approx 2$ would
bring $\epe$ in agreement with the measured value for central values of the
other parameters. Let us further discuss possible scenarios within the
Standard Model which would yield consistency of $\epe$ with the experimental
measurements without requiring additional new physics contributions.

The calculations of the $B$-parameters $B_6^{(1/2)}$ and $B_8^{(3/2)}$
involve non-perturbative physics and are thus still very uncertain. Whereas
the value of $B_8^{(3/2)}$ seems to be under better theoretical control,
it could well be that the ranges as given in table~\ref{input} underestimate
$B_6^{(1/2)}$. Hints are given by the large correction of ${\cal O}(p^2/N_c)$
in the large-$N_c$ approach. Additional indications in this direction come
from the recent work in refs.~\cite{bp:95}. Also using large-$N_c$ methods
and an intermediate colour-singlet boson which is claimed to provide the
correct matching between the short-distance Wilson coefficients and the
hadronic matrix elements, the authors of \cite{bp:95} obtain
$B_6^{(1/2)}=2.2\pm 0.5$. Although premature at this stage the result is
certainly interesting as it would provide the required enhancement of $\epe$.

Another contribution to $\epe$ which deserves to be reconsidered are the
isospin-breaking corrections. The original calculations \cite{omIB1,omIB2}
which more than ten years ago estimated $\Omega_{IB}\approx 0.25$, only
considered $\pi^0$-$\eta$, $\eta'$ mixing as the source for isospin-breaking.
As pointed out in the recent work \cite{gv:99}, additional isospin-violating
effects arise from the $u$-$d$ quark mass difference directly. Estimating
these additional contributions in chiral perturbation theory with resonances
\cite{egpr:89} the authors of \cite{gv:99} find that $\Omega_{IB}$ might
even change sign and become as low as $\Omega_{IB}\approx -\,0.6$, depending
on the couplings of the scalar resonance sector. Such a change effectively
would correspond to $B_6^{(1/2)}\approx 2$, again bringing $\epe$ in
agreement with the experimental average. However, in this case the couplings
of the scalar resonances are rather uncertain and the findings in
ref.~\cite{gv:99} need further corroboration.

The final point that should be discussed here is the issue of final state
interactions. In principle, non-perturbative approaches to the hadronic
matrix elements should also reproduce the strong final-state phases of
the $\pi\pi$ system. At present, however, since these phases are generated
by chiral loops, in all approaches to the matrix elements they are either
zero \cite{bbg:87}, or found substantially smaller than the experimental
values \cite{BNc,BChQM}. A first step in the direction to fully include
final-state interaction effects in the calculation of $\epe$ has been
taken very recently in ref.~\cite{pp:99}.

In the elastic region for the $\pi\pi$ scattering, unitarity and analyticity
constraints permit to give a representation of the isospin amplitudes $A_I$
in terms of the so-called Omn{\`e}s integral \cite{omn:58} which involves
the phase shifts $\delta_I$, times an arbitrary polynomial in momenta. Thus,
the effects of chiral logarithms are resummed to all orders in the Omn{\`e}s
integral and the polynomial ambiguity can in principal be fixed by a
calculation in chiral perturbation theory. Taking these steps with lowest
order chiral expressions for the polynomial ambiguity, the authors of
\cite{pp:99} found that besides the imaginary parts which reproduce the
final state phases by construction, there is also a substantial enhancement
of the real part for isospin zero and a slight suppression for isospin two.
The corresponding enhancement and suppression factors $\Re_I$ were
estimated to be
\begin{equation}
\label{FSI}
\Re_0 = 1.41 \pm 0.06 \,, \quad
\Re_2 = 0.92 \pm 0.02 \,,
\end{equation}
respectively. Applying these factors in the expression for $\epe$, for
central values of the parameters $\epe=15\cdot 10^{-4}$, much closer to
the experimental result.

As they stand, the results by the authors of ref. \cite{pp:99} are very
interesting and may provide the dominant source of enhancement required
to bring theoretical calculations of $\epe$ within the Standard Model
and the experimental results into agreement. These findings might also
be linked to large values of $B_6^{(1/2)}$ found in the large-$N_c$
approach. Nevertheless, it would be important to demonstrate that the
factors $\Re_I$ take the values of equation \eqn{FSI} also in the calculation
of individual matrix elements in particular approaches like large-$N_c$
or the chiral quark model.

Even though from the discussion above it appears that new-physics
contributions at present are not required to fit the data for $\epe$,
there is certainly still room for such contributions. The most plausible
sizable contribution could come from chromo-magnetic penguins in general
supersymmetric models or modified $Z$-penguins. On the other hand substantial
modifications of QCD penguins through new physics are rather improbable.
For a further discussion of contributions to $\epe$ beyond the Standard
Model, the reader is referred to the talk by Masiero \cite{mas:99}.

The future of $\epe$ in the Standard Model and its extensions will depend
strongly on the progress which is reached in the calculation of hadronic
matrix elements. This progress should include a reliable calculation of
$B_6^{(1/2)}$ on the lattice, control over the scale and scheme dependencies
in approaches using effective low energy theories such that a proper
matching with the Wilson coefficient functions at the next-to-leading order
will be possible, a better understanding of isospin-breaking effects,
and finally, the proper inclusion of final state interactions. First
successes in all these areas have been achieved and, together with the
upcoming improvements of the experimental measurements with increased
data sets, to my mind the future for $\epe$ in the new millennium looks
bright.

\acknowledgments
It is a pleasure to thank my collaborators for a most enjoyable time
working on $\epe$ and the organisers for a very pleasant and interesting
meeting on heavy flavour physics. I would also like to thank the
Deutsche Forschungsgemeinschaft for their support.

\end{document}